# Observation of quantum transport features in graphene devices fabricated utilizing a nano-manipulating probe technique

Christopher Coleman, Davie Mtsuko, Chris Botha[†] and Somnath Bhattacharyya[a]

Nano-Scale Transport Physics Laboratory, School of Physics, and DST/NRF Centre of Excellence in Strong Materials, University of the Witwatersrand, Johannesburg, Private Bag 3, WITS 2050, South Africa

[†] Present address: Orbilectron Consulting (Pty) Ltd., Port Elizabeth, South Africa

A novel method for fast fabrication of mesoscopic multilayered graphene electronic devices utilizing nanoprobes to exfoliate graphite flakes is developed. The magnetoresistance of these devices exhibit pronounced Shubnikov-de Haas oscillations at magnetic fields above 4 T and at temperatures below 30 K. From the analysis of the SdH oscillations we show that multilayer graphene devices have a carrier density and effective mass ($m^* = 0.042 m_e$ - $0.083 m_e$) comparable to those of bilayer and trilayer graphene. The quantum lifetime in this multilayered graphene is in the range 22 to 90 fs corresponding to a disorder-broadening of 5 to 15 meV.

[a] Author to whom correspondence should be addressed. Electronic mail: Somnath.Bhattacharyya@wits.ac.za

**Introduction:**

Graphene, the two-dimensional allotrope of carbon has since its discovery in 2004 attracted an immense amount of scientific attention [1-3]. This is due to the unique and fascinating properties of this material such as high current densities of orders of magnitude larger than in copper and also the existence of Dirac fermions at the *k*-points [4,5]. Since this discovery, graphene has been hailed as a super material with the possibility of replacing silicon for electronic purposes and a potential material in the field of micro and nano-electronics. There are however many challenges in creating functional devices with graphene, these range from the isolation of single or few layered graphene sheets to the transfer and manipulation of the material for device fabrication. Conventional methods used in the fabrication of these devices include delicate and complex processes such as electron beam lithography coupled with reactive plasma etching as well as techniques using scanning probe microscopy specifically AFM-lithography [1,6,7]. In addition to being relatively costly these processes have proved to be time consuming with an extremely low output of functioning devices, leading to a search for more expeditious and effective routes for the device fabrication. In this work we develop a novel method of fast and contamination-free fabrication of few layer graphene devices. This involves mechanically peeling off thin sheets of multilayered graphene from bulk exfoliated graphite stock, placement of the multilayers onto prefabricated Hall bar and waveguide contact pads, in order to create devices for Hall effect and magnetoresistance (MR) measurements as well as high frequency measurements. By comparison with results obtained in single layer, bilayer and trilayer graphene a strong understanding of the origin of the evolution of transport characteristics in the transition from single layer to graphite can be established. Magnetic field-dependent studies on single layer graphene have shown Shubnikov de Haas oscillations which get damped as the temperature increases or the bias current is increased as well as carrier lifetimes up to 53 fs and effective masses $0.033m_e$ [8].

The influence of interlayer effects on SdH oscillations and transport parameters in multilayer graphene devices is of great interest. Temperature damped SdH measurements on bilayer graphene in one study have found hole and electron effective masses of $0.034m_e$ and $0.046m_e$, respectively, with the suppressed $m^*$ values (with respect to single particle values) being attributed to a renormalization of band structure induced by electron-electron interaction [9]. In other studies on trilayer graphene effective mass of $0.054m_e$ were obtained showing that effective mass increases with increasing number of layers [10]. It is of interest to see if this trend continues as the number of layers is further increased. In works on multilayered graphene systems it was also established that the electric field effect is prominent and that the frequency of the SdH oscillations varies with a change in the gate voltage of the devices [11-13]. Quantum Hall effect studies in ABC stacked trilayer graphene devices, have revealed a new type of quasiparticle with $l=3$ chiral degree and a cubic dispersion predicted by theoretical works [14]. Despite the complexity of multilayer graphene systems such a richness in transport physics observed in bilayer and trilayer is a strong drive of our studies of multilayer systems.

**Experimental details:**

The multilayered graphene stock samples were obtained by the conventional scotch tape method developed by Konstantin Novoselov and Andre Geim whereby highly ordered pyrolytic graphite is continually mechanically cleaved using an adhesive tape until few layers remain [1]. These layers are visible under an optical microscope even at low magnification with different thicknesses or number of layers rather, appearing as different colored flakes due to interference effects. This method is usually continued until one arrives at single layer graphene. As can be imagined this method is relatively time consuming. Also other complex e-beam lithography procedures (including mapping, resist coating, write field alignment, additional metallization and lift-off steps) need to be carried out in order to produce devices for the electrical characterization. In our approach, the multilayered graphene samples on the SiO$_2$ substrates are annealed at a temperature of 1000 $^{\circ}$C in argon gas environment, then imaged using a JEOL J1007F SEM.

Using nano-manipulators (Kleindek-Germany), with tips having a radius of curvature of approximately 100 nm, thin strips of layered graphene were peeled off of the larger flakes. It was found that these strips adhered relatively well to the manipulating probe and once separated from the larger flakes were easily transported by moving the probe to any desired location within the imaging range. It was found, however, that placing the graphene strips down onto the desired location was difficult due to substrate charging, this was overcome by grounding the substrate in the SEM. In this case multilayered graphene strips were placed and positioned onto prefabricated contact pads designed in six-probe Hall bar configuration. It was observed that using this method introduces ripples and folds in the material, this is unwanted as it would affect the transport properties of the device. Attempts were made at using the nano-probe tips to iron out these ripples, this technique showed some success with larger wider ripples but was unable to smooth out sharp thin folds in the multilayered sheet. The multilayered strips where then welded into place with platinum tabs using a combination of SEM beam and gas injection system (OmniProbe USA). Once the multilayered strips were secured, the devices were annealed in argon to remove any contaminants and then subjected to Raman analysis. Although the described process has yielded only multilayered sheets of graphene thus far, the technique is far less time consuming and provides a mechanism to suspend sheets directly across contact electrodes. Utilizing this method also eliminates much of the risk of contaminating and functionalizing the graphene samples with chemicals commonly used in device fabrication.

Fig. 2 shows a Raman spectrum for the multilayered graphene device created using the above method. The prominent features of the multilayered sample used in this study include the 2D peak at 2700 cm$^{-1}$ which has a lower intensity than the G peak at 1580 cm$^{-1}$. This is unlike single layer graphene which has a 2D peak with much greater intensity than that of the G peak. The shape of the 2D band is routinely used to determine the number of layers in few layered samples. This is possible due to the unique peak shapes which can be fitted with multiple Lorentzians specific to the number of layers up to 4 layers, beyond this number the 2D peak is too similar to that of graphite to determine number of layers. The sample used in this study shows a 2D peak that can be fitted with two Lorentzian curves, the one at higher frequency being more prominent, as is expected for graphite. This implies that the sample is composed of more than 4 layers as the unique spectra are only observed up to this point. Also noted is the small D peak at 1350 cm$^{-1}$. This Raman peak is an indication of disorder in the material. The introduction of disorder is mostly due to folding and rippling of the sheet layers which was found to be unavoidable.

**Results and Discussion:**

In bench-top measurements the resistance was found to be in the Ω-range. This is expected as having a multilayered system allows for multiple conduction channels, increasing the conductivity compared to single or very few layered graphene devices where resistance is typically in the kΩ range. The devices were then cooled to room temperature and characterized over a large temperature range. The Hall bar device was analyzed using a cryogenic cryostat system with the capabilities of achieving measurements in fields up to 12 T. The resistance-temperature dependence at various fields, perpendicular to the sample surface, is shown in Fig. 3(a). It is noted that initially the device exhibits a decrease in resistance as temperature is raised, but only up until 25K, at this point the resistance increases with temperature. This change indicates a crossover of conduction mechanism from insulating to metallic behavior. Also noted are resistance fluctuations which occur at high fields above 4T and temperatures below 35 K. To investigate this oscillatory behavior the MR measurement of the device was undertaken (See Fig. 3(b)). Here it was found that at low temperatures below 25 K, initially the device shows a negative MR in the low field regime (from 0 T to 4 T), above 4 T however the MR starts to oscillate with increasing amplitudes the magnetic field is ramped up. These oscillations are noted to be damped as the temperature is increased. As the magnetic field is increased the Fermi sphere splits into Landau tubes. The change is temperature $k_B \Delta T$ shifts the Landau levels with respect to the Fermi level producing resonances when the Fermi level coincides with the Landau levels. As temperature phonon-induced disorder broadening smears the oscillations.

This oscillatory behavior persists until 35 K, above this temperature no oscillations were observed. This is consistent with the interpretation of the *R-T* data as well as previous studies that have shown that SdH oscillations are observable in mesoscopic graphite and thin layered graphene [1,8]. From this MR data, the resistance as a function of inverse magnetic field was plotted at temperatures 5 K, 10 K, 15 K and 20 K. By determining the SdH frequency it is able to calculate carrier density for the device by the equation: $v = \frac{n_e h}{eB}$. Here, $n_e$ is the carrier density, *B* the applied magnetic field with *h* and *e* being Planck's constant and the fundamental charge. The carrier density was found to decrease as temperature increased and also to increase slightly with increasing magnetic field. There were two exceptions to this general trend; the MR measurements made at 2K and at 5K. At these low temperatures it was expected that SdH oscillations would be most prominent, that however was found not to be the case. At these two lowest temperatures oscillatory behavior did start at 4T applied field but the strength of the oscillations are miniscule as compared to that of the oscillations at higher temperatures.

From the carrier density we can evaluate the effective mass $m^* = (\frac{\hbar}{v_F})\sqrt{\pi n_e}$. For carrier density in the range $5 \times 10^{12}$cm$^{-2}$ to $2 \times 10^{13}$ we find the mean effective mass of in our multilayer graphene device to be in the range $0.042m_e$ to $0.083m_e$. This is comparable with the effective mass measured in bilayer layer and trilayer graphene and reported in literature [9] and the theoretical values of effective mass of AB stacked bilayer graphene ($m_{AB} \approx \gamma_1/2v^2 \approx 0.035 m_e$) and ABA stacked trilayer graphene ($m_{ABA} = \sqrt{2} m_{AB} \approx 0.05 m_e$) [14]. Our measurements confirm that the increase in number of layers leads to higher effective mass compared to effective mass of monolayer graphene.

Apart from the carrier density shown in Fig. 3(c), we can extract two extra transport parameters from the SdH oscillation data analysis namely the carrier lifetime or the quantum scattering time and the phase coherence length. To extract the quantum lifetime we make use of the general form of MR of a two-dimensional electron system (2DES) in the SdH oscillation regime which is given by [8]: $R_{XX} = R_0 \left[ 1 + \lambda \sum_{S=1}^{\infty} D(sX) exp\left(-\frac{s\pi}{w_c \tau}\right) \times \cos\left(s\frac{\hbar S_F}{eB} - s\pi + s\phi_0\right) \right]$, from which to the first order we can write the amplitude of oscillations as: $A_{ex} = \lambda D(X) exp\left(-\frac{\pi}{w_c \tau}\right)$, where $w_c = eB/m^*$ is the cyclotron frequency and $D(X) = \frac{2\pi k_B T/\hbar w_c}{sinh(2\pi k_B T/\hbar w_c)}$ is the damping factor. Hence by plotting the amplitudes ($A_{ex}/D(X)$) versus 1/$B_{ex}$ of the oscillation at extrema as shown in Fig. 1, we find the carrier lifetime in the range 22 fs to 90 fs for effective mass 0.065 $m_e$. This is comparable to a lifetime of 54 fs observed in single layer graphene [8]. In Fig. 2 we plot the lifetime as a function of temperature. The $\tau$ range corresponds to a disorder-broadening values ($\Gamma = \hbar/2\tau$) in the range 4.6 to 15 meV. This is comparable to disorder broadening (5.5 to 8.0 meV) observed in bilayer graphene [9]. The phase coherence length is estimated using formula $\Delta B = \varphi_0/L_\varphi^2$ = to be 66 nm at 25 K at 6 T and 75 nm at 6 T at 2 K. [15].

The inverse period of the SdH oscillation can also be used to estimate the cross sectional area of the Fermi surface pocket perpendicular to the magnetic field using the formula $\Delta(1/B) = 2\pi e/\hbar A$ [16,17]. For a 10 T field at 25 K we find the extremal Fermi surface area to be ~9×10$^{17}$ m$^{-2}$. However, the exact morphology of the Fermi surface pockets is not known.

We next examine some peculiar features of the SdH oscillations. It is seen that as the temperature is increased the amplitude of oscillations in the high field regime slightly reduces. This damping of oscillation is due to phonon scattering which reduces the phase coherence length. It is also observed that the extrema shift location on the $1/B$ axis as the temperature is varied. This can be explained by small shifts in Landau level positions due to inelastic phonon scattering.

We note the manifestation of beat frequencies in the MR with the field arranged perpendicular to the sample. The formation of these beat frequencies is generally understood to indicate coherent transport between layers whereby electrons are transferred directly between layers ensuring that intralayer momentum is conserved. As a consequence interference between wavefunctions on adjacent layers is possible [18]. It is also possible for this mode of interlayer transport to be weakly incoherent as the tunneling events between layers are uncorrelated due to the electrons being scattered multiple times within the layers before tunneling between the layers. This beat behavior is related to the warping of the Fermi surface and the interference of cyclotron orbits of different radii, the so called short and long orbits [18]. Furthermore it has been experimentally demonstrated that transport between few layer graphene is strongly related to the stacking order of the layers and breakdown of coherent transport can occur due to twisting or misalignment whereby the stacking order is changed from the Bernal to rhombohedral configuration [19]. Taking all these aspects into consideration it is quite natural to link the modulated SdH oscillations to these interlayer interference effects. The irregularity in the oscillations may also arise from the modulation of the SdHO by two other kinds of oscillations known as magneto-intersubband oscillations (MISO) and phonon induced resistance oscillations whose origin can be traced to scattering-assisted coupling of carrier states in different Landau levels [20].

**Conclusion:**

In summary, we have demonstrated a quick and efficient way of producing multilayer to a few layer graphene devices. The proposed method is faster than conventional procedures and eliminates much of the risk for contamination but has the disadvantage of being an extremely delicate process. This method does not allow for the peeling off of single layer graphene owing to the large tip radius of the manipulating probe. It has however proven to be effective at creating graphene devices in the mesoscopic regime such that it can be regarded to hold much potential for fast device fabrication for research purposes in the field of mesoscopic layered systems. The low temperature measurement exhibit quantum effects in the form of SdH oscillations. From the analysis of the SdH oscillations it is seen that this multilayer graphene device has carrier density, effective mass, quantum lifetime and phase coherence length comparable to those of bilayer and trilayer graphene devices.

*Acknowledgements:* SB acknowledges the CSIR-NLC for establishing the laser ablation set up which produced the some of the graphene samples and NRF (SA) NNEP as well as nanotechnology flagship grant.

**Figure Captions:**

**Fig. 1.** SEM images of the main steps followed in the proposed method in creating multilayered devices. **(a)** Thin strips of layered graphene are peeled off of larger flakes. **(b)** The strips are transported to desired location using the probe tip. **(c)** Gas injection is used to make platinum tabs to keep strips in place. **(d)** The final device ready for a four or six probe measurement.

**Fig. 2.** Raman spectrum of multilayered graphene. **Inset:** Decomposition of 2D peak.

**Fig. 3. (a)** Resistance temperature dependence of four-probe device at various magnetic fields (0 T, 4 T, 8 T, and 12 T). **(b)** The resistance versus positive applied field for temperatures 15 K and 20 K showing oscillation peaks. **(c)** Calculated carrier density as a function of applied magnetic field at different temperatures. **(d)** A plot of the resistance versus applied positive field for temperatures 2K, 15 K and 25 K.

**Fig. 4.** A comparison of the longitudinal resistance as a function of inverse applied magnetic field. Although oscillations do not occur at the same field there is a correspondence in the period at different temperatures **(a)** 25 K, **(b)** 20 K, **(c)** 15 K, and **(d)** 10 K.

**Fig. 5. (a)** A plot of oscillation extrema amplitude ($A_{ex}$) versus inverse of magnetic field. **(b)** The variation of quantum scattering time with temperature.

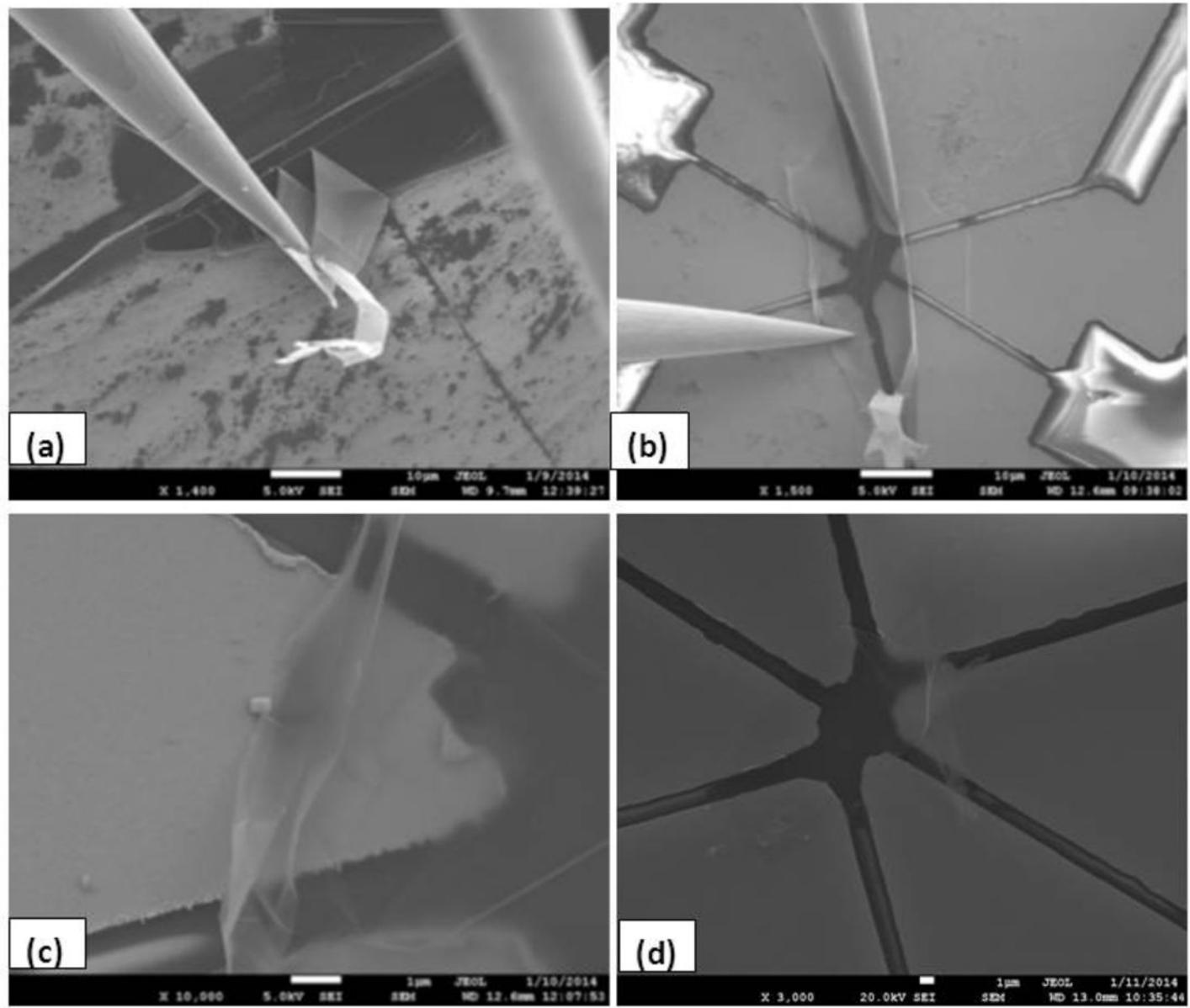

Fig. 1 *(a) - (d)*

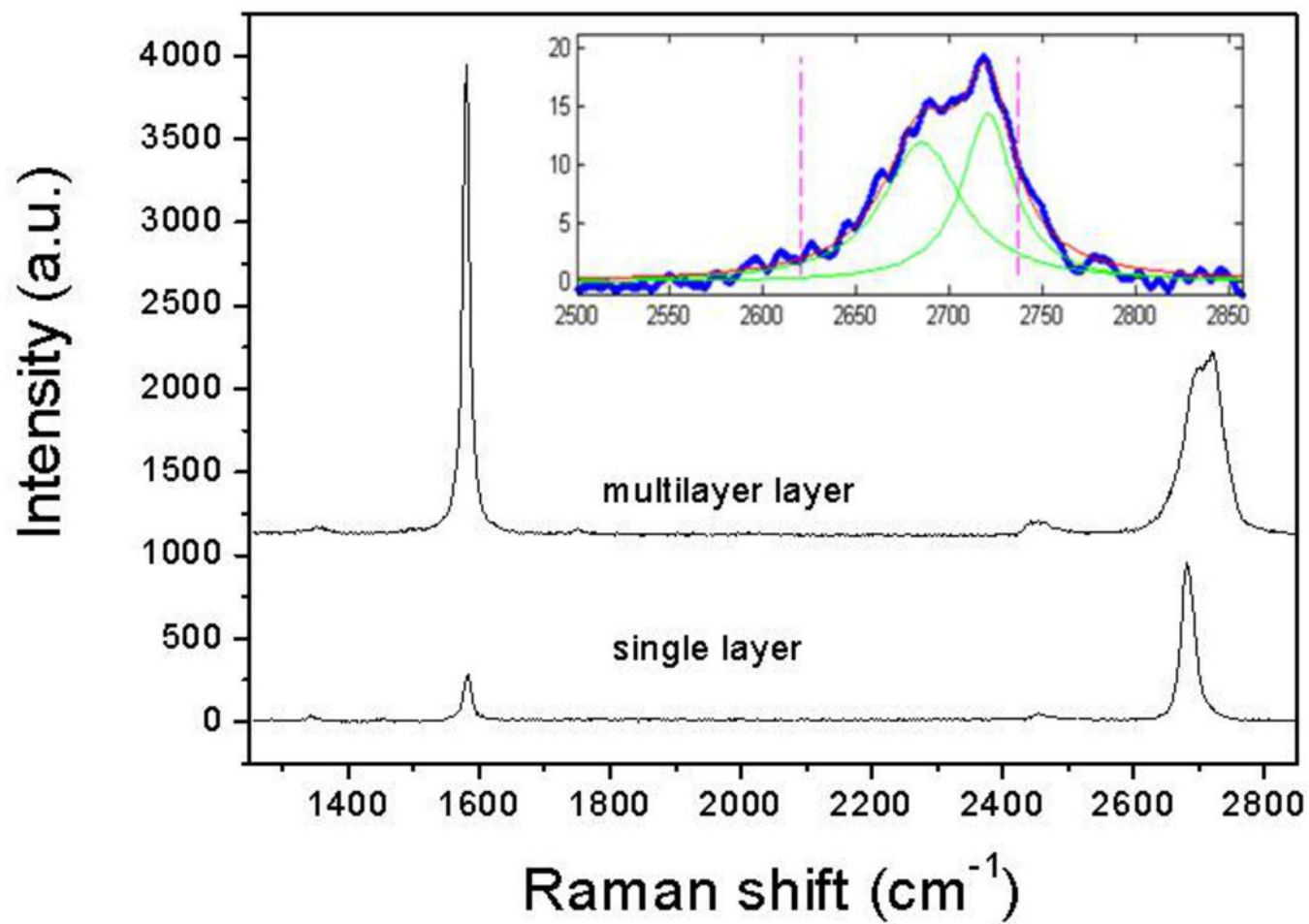

Fig. 2

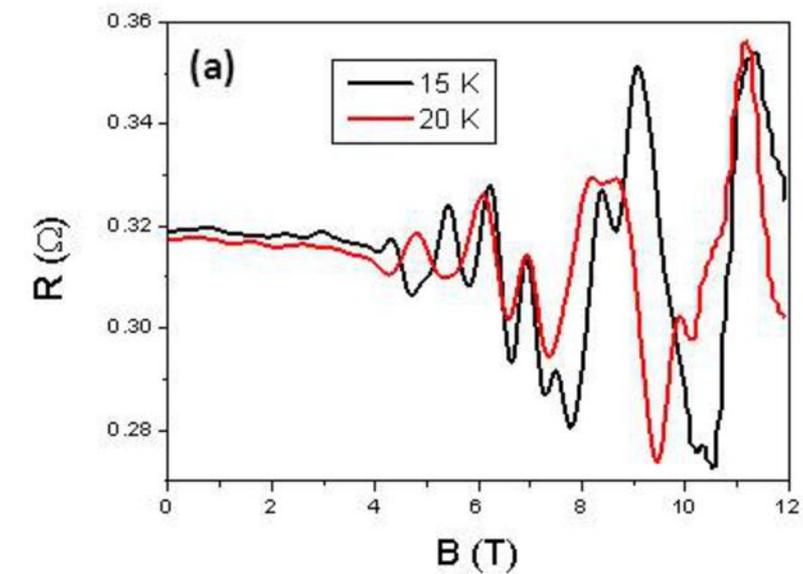
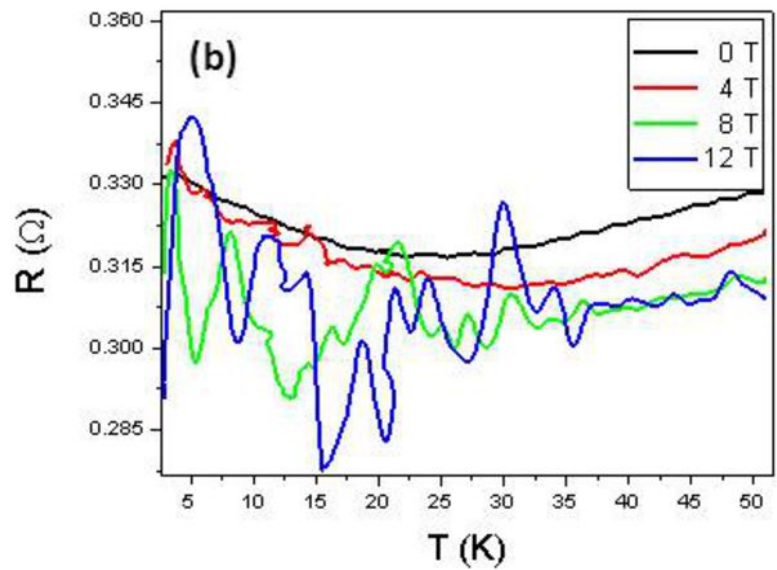
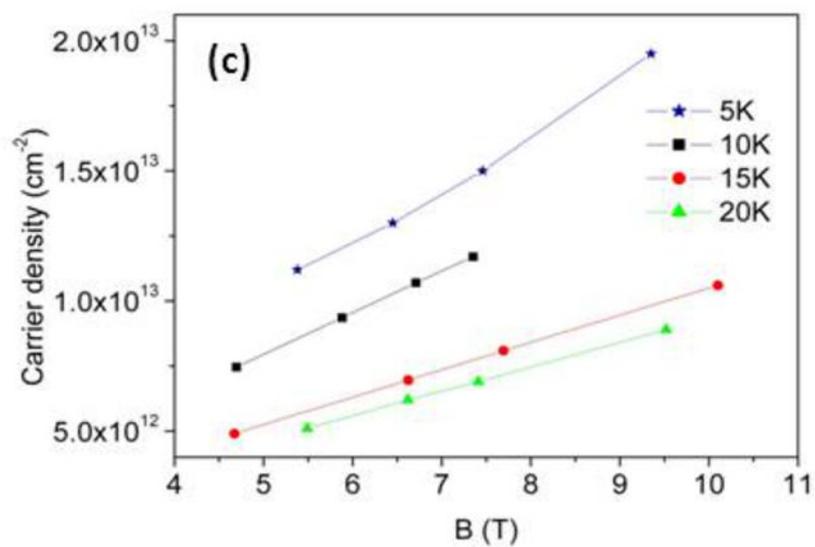
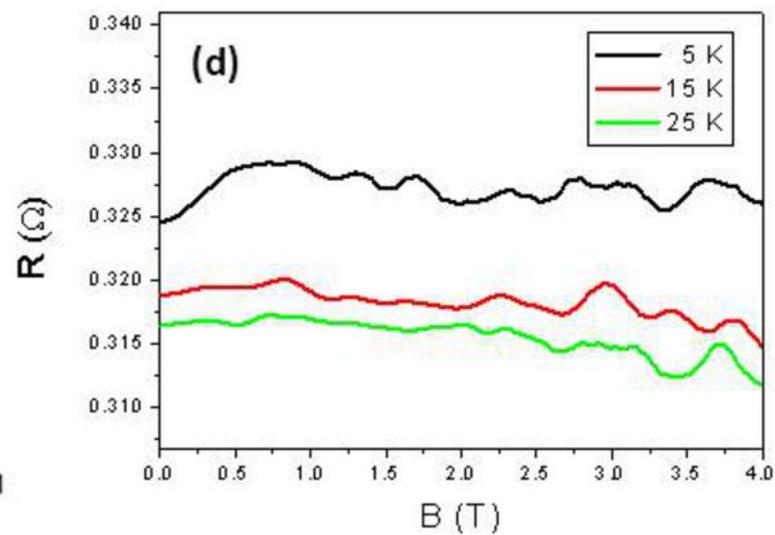

Fig. 3 *(a) - (d)*

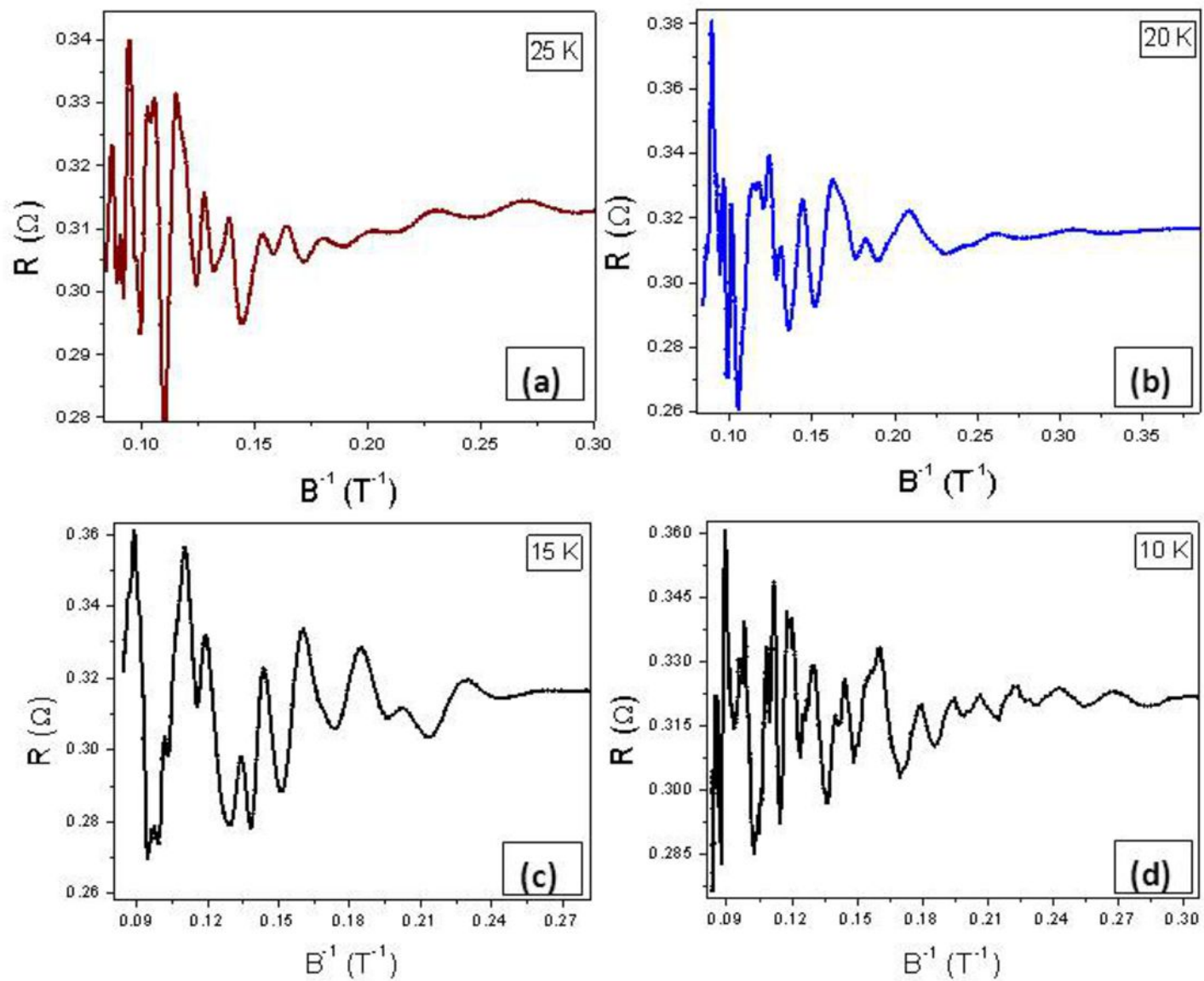

Fig. 4 *(a) - (d)*

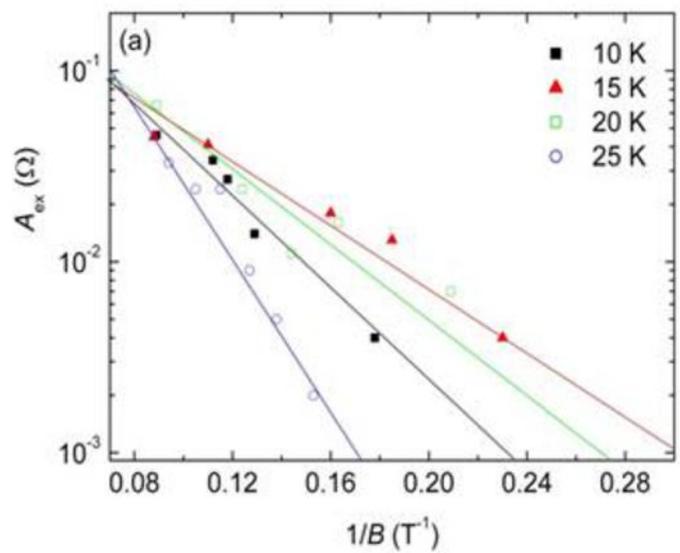
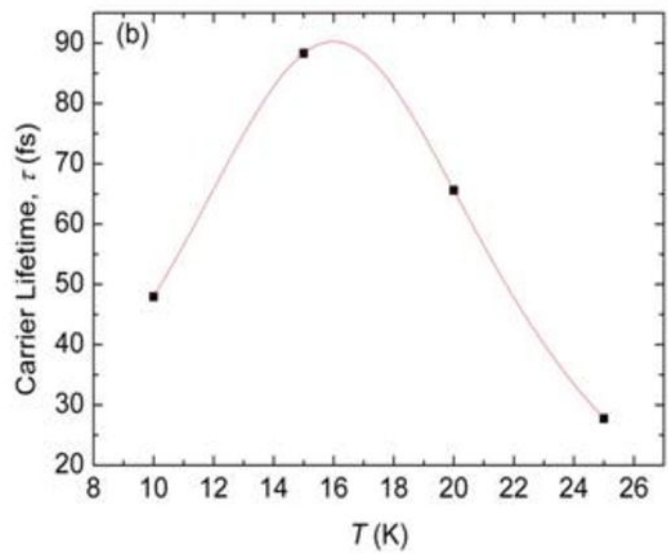

Fig. 5 *(a) and (b)*